\documentclass[a4paper, preprint, preprintnumbers, superscriptaddress, showpacs, showkeys, nofootinbib, floatfix, twocolumn, aps, pra, 10pt]{revtex4-1}
\usepackage[utf8]{inputenc}
\usepackage[english]{babel}
\usepackage[dvips]{graphicx}
\usepackage[breaklinks=true,colorlinks,citecolor=blue,linkcolor=blue,urlcolor=blue]{hyperref}
\usepackage{xcolor,url,amsthm,amsmath,amssymb,amsmath,
mathtools,latexsym,appendix,physics,bbm,enumerate}

\usepackage{tikz,tkz-euclide}
\usetikzlibrary{shapes, calc, arrows, positioning, through,scopes, angles, quotes, fit, shadings, arrows.meta}

\usepackage{orcidlink}

\newtheorem{theorem}{Theorem}

\allowdisplaybreaks

\begin{document}

\title[LAQC Swap]{Local-available quantum correlation swapping in one-parameter X states}

\author{Hermann L. Albrecht\,\orcidlink{0000-0002-5735-8340}}
\email{albrecht@usb.ve}
\affiliation{Departamento de F\'{\i}sica, Universidad Sim\'on Bol\'{\i}var, AP 89000, Caracas 1080, Venezuela.}

\date{December 29, 2025}% It is always \today, today,
             %  but any date may be explicitly specified

\begin{abstract}
Although introduced for entanglement, quantum repeaters and swapping protocols have been analyzed for other quantum correlations (QC), such as quantum discord. Introduced by Mundarain and Ladrón de Guevara [Quantum Inf. Process.
14, 4493 (2015)], local-available quantum correlations (LAQC) are a promising yet understudied quantum correlation. Recently, Bellorin et al. [Int. J.
Mod. Phys. B 36, 22500990 (2022), Int. J. Mod. Phys. B 36, 2250154 (2022)] obtained exact analytical results for the LAQC quantifier of general 2-qubit X states. Building up from those results, we analyzed the LAQC swapping for 2-qubit X states. As expected, we find that if the initial states are non-classical and the one used for the projective measurement is entangled, the final state will generally have non-zero LAQC. Using the properties of this quantum correlation, we establish the conditions for a QCS scheme that leads to a final state with a non-zero LAQC measure. We illustrate these results by analyzing five families of one-parameter 2-qubit X states, including families where the projective measure leads to a separable state, but whose LAQC measure is non-zero. This feature opens the possibility for this quantum correlation to be considered a genuine resource in quantum information technology.
\end{abstract}
\preprint{SB/F/497-25}
\maketitle

%%%%%%%%%%%%%%%%%%%%%%%%%%%%%%%%%%%%%%%%%%%%%%
%%%%%%%%%%%%%%%%%%%%%%%%%%%%%%%%%%%%%%%%%%%%%%
%%%%%%%%%%%%% Introducción %%%%%%%%%%%%%%%%%%%
%%%%%%%%%%%%%%%%%%%%%%%%%%%%%%%%%%%%%%%%%%%%%%
%%%%%%%%%%%%%%%%%%%%%%%%%%%%%%%%%%%%%%%%%%%%%%

Since quantum correlations are the key ingredient for quantum advantage and quantum supremacy in quantum communications and computation, their quantification, control, and manipulation are central tasks in quantum information science and technology. Whether focused on a local quantum computer \cite{Nakahara-QC, Nielsen-QIT} or aiming for the quantum internet \cite{Davis2025EntSwaoQuantumInternet}, the study of quantum networks \cite{Chiribella2009TheoreticalQNetwork, Nokkala2024ComplexQNetwork-Review,Azuma2023QuantumNetwork-Review} and the manipulation and redistribution of quantum correlations in them is an active field of research, both theoretically and experimentally.

Entanglement \cite{Horodecki-Ent}  is still the preferred quantum correlation of interest. Since the introduction of quantum discord \cite{qDiscord-Olliver, qDiscord-Henderson}, other quantum correlations have been defined and studied \cite{Modi-qDiscord}. The ability of a local measurement to affect the global state depends on whether it is correlated or not. This intrinsic characteristic of quantum mechanics, deemed by Einstein as ``\emph{spooky action-at-a-distance}" \cite{EPR-1935}, is at the core of the development of numerous protocols and applications in quantum communications \cite{Bennett1993Teleporting, Vaidman1994Teleportation}, cryptography \cite{Gisin2002QuantumCrypto, Pirandola2020AdvancesCrypto}, and more.

Although most have relied mainly on entanglement as the quantum resource, other  introduced correlations have also found a place in practical applications. For instance, several studies have analyzed the operational interpretation \cite{Cavalcanti2011QDOperInterpret} and significance \cite{Gu2012QD-OperatSignif} of quantum discord, while others have focused on its role and importance in remote state preparation \cite{Dakic2012-QD-RemoteStatePrep}, quantum communication \cite{MADHOK2012QD-QuantumComm, Datta2017QDCommProtocols}, and quantum cryptography \cite{Pirandola2014QDQuantumCrypto}, among others.

Entanglement swapping \cite{Yurke1992EPR-NonInteract, PhysRevA1992YurkeStoler, Zukowski1993-EntSwap}, a scheme closely related to quantum teleportation, is crucial to redistribute entanglement in a quantum network. As such, it is one of the tools to implement quantum repeaters \cite{Azuma2023QuantumNetwork-Review, Wallnöfer2024QuantumRepeater}. Since its proposal, experimenters have been able to implement entanglement swapping of qubit systems, and several authors have generalized the protocol to involve higher-dimensional and continuous variable systems \cite{Pan1998Exp-EntSwap, Jennewein2001Exp-EntSwap, PRA2002-ExpEntSwap-NMR, PhysRevA2003-ExpEntSwap, Xiao2006-ExpEntSwap-ChinPhys, Riebe2008DetermEntSwap-Exp, Schmid2009-ExpEntSwap-Optics, PhysRevA2014-ExpEntSwap-Matter}. Recently, Cai et al. published a review \cite{Cai2025-QCsharing-review} on quantum correlation sharing and its applications, exploring examples such as  quantum random access codes, random number generation, and self-testing tasks.

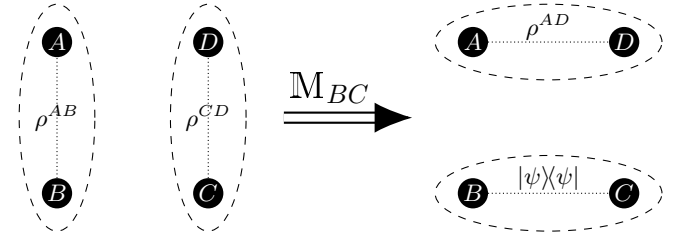
\begin{figure}
\begin{center}
\begin{tikzpicture}
    %Gradilla de Ayuda
    %\draw[help lines] (0,0) grid (8,3);
    %rho^{AB}
    \fill (0,2) circle(2mm);
    \node at (0,2) {\textcolor{white}{$A$}};
    \fill (0,0) circle(2mm);
    \node at (0,0) {\textcolor{white}{$B$}};
    \draw[densely dotted] (0,0.2)--(0,1.8);
    \draw[dashed] (0,1) ellipse (5mm and 15mm);
    \node at (0,1) {$\rho^{AB}$};
    %rho^{CD}
    \fill (2,0) circle(2mm);
    \node at (2,0) {\textcolor{white}{$C$}};
    \fill (2,2) circle(2mm);
    \node at (2,2) {\textcolor{white}{$D$}};
    \draw[densely dotted] (2,0.2)--(2,1.8);
    \draw[dashed] (2,1) ellipse (5mm and 15mm);
    \node at (2,1) {$\rho^{CD}$};
    %Flecha Medición
    \draw[thick,double,double distance=3pt,->,>={Latex[length=5mm, width=4mm]}] (3,1)--(4.7,1);
    \node at (3.6,1.4) {\Large $\mathbbm{M}_{BC}$};
    %rho^{AD}
    \fill (5.5,2) circle(2mm);
    \node at (5.5,2) {\textcolor{white}{$A$}};
   \fill (7.5,2) circle(2mm);
    \node at (7.5,2) {\textcolor{white}{$D$}};
    \draw[densely dotted] (5.7,2)--(7.3,2);
    \draw[dashed] (6.5,2) ellipse (15mm and 5mm);
    \node at (6.5,2.2) {$\rho^{AD}$};
    %rho^{BC}
    \fill (5.5,0) circle(2mm);
    \node at (5.5,0) {\textcolor{white}{$B$}};
    \fill (7.5,0) circle(2mm);
    \node at (7.5,0) {\textcolor{white}{$C$}};
    \draw[densely dotted] (5.7,0)--(7.3,0);
    \draw[dashed] (6.5,0) ellipse (15mm and 5mm);
    \node at (6.5,.2) {$\dyad{\psi}$};
\end{tikzpicture}
    \caption{Schematic representation of quantum correlation swapping via a projective measurement using a pure state $\ket{\psi}$.}
    \label{fig:QCSwappingProtocol}
\end{center}
\end{figure}

By applying a projective measurement using a maximally entangled state, such as a Bell state, two qubits that have not interacted can become entangled. Consider two pairs of bipartite systems, characterized by the density operators $\rho^{AB}$ and $\rho^{CD}$. Since these states are not interacting, one can write the global state as $\rho^{ABCD}=\rho^{AB}\otimes\rho^{CD}$. If one applies a projective measurement $\mathbbm{M}=\dyad{\psi_{ME}}$ on subsystems $B$ and $C$, the resulting state is given by $\tilde{\rho}^{ABCD} = \rho^{AD}\otimes\dyad{\psi^{BC}_{ME}}$. We present a schematic illustration of a quantum correlation swapping protocol via a projective (von Neumann) measurement in Figure \ref{fig:QCSwappingProtocol}.

Although the entanglement swapping protocol allows one to redistribute this quantum correlation to non-interacting subsystems, it comes with a cost. The maximal amount of entanglement that one can redistribute depends on the initial states and the projective measurement involved. In 2014, Muñoz, Grüning, and Roa  \cite{Roa2014EntSwap-Xstates} analyzed this scheme for X states, obtaining threshold values for the initial states’ entanglement so that the outcome states are not separable.

In 2015, Xie et al. \cite{Xie2014QCSwapping} extended the concept of entanglement swapping to deal with other quantum correlations. Therefore, they deemed their proposal as Quantum Correlation Swapping (QCS). Using the same basic protocol described above, the authors analyzed the redistribution of quantum discord, measurement-induced disturbance \cite{Luo_MID}, and its ameliorated version \cite{Girolami_AMID, Wu_AMID}.They used Werner states as their initial 2-qubit states in their first paper, and later extended their analysis to include Bell-Diagonal states \cite{Xie2019QCSwapping}. In it, the authors focused on analyzing quantum discord and local quantum uncertainty \cite{Girolami2013LocalQuantumUncertainty} as the quantum correlations they are swapping. More recently, Xie et al. \cite{Xie2025Quantum} have extended their analysis to include X states.

Since we can apply the QCS scheme to various quantum correlations, we analyze the behavior of local-available quantum correlations (LAQC) \cite{LAQC}. Introduced in 2015 by Mundarain and Ladrón de Guevara, LAQC has yet to become a more commonly studied quantum resource, despite recent results showing them to be more robust than entanglement when subjected to Markovian decoherence \cite{LAQC_BD, LAQC_BD-Err}. Contrary to what one can achieve for quantum discord, where only approximate analytical results exist, an exact result for the LAQC quantifier for general 2-qubit X states was derived \cite{LAQC_Xstates-sym,LAQC_Xstates-no_sym}. Therefore, we have focused on studying LAQC swapping by considering relevant one-parameter 2-qubit X states as initial states.

We have organized this work as follows. In Section  \ref{sec:Xstates}, we introduce 2-qubit X states and their properties.Then, w proceed to present five relevant families of one-parameter X states, namely Werner states \cite{Werner}, $\alpha$ and $\beta$-states \cite{Qasimi2011-AlphaBeta-Xstates}, one-parameter mixed states involving a basis element \cite{Verstraete-MAF}, and maximally entangled mixed states \cite{Munro2001Maximizing}. Next, we review the definition of local-available quantum correlations and the results obtained for general 2-qubit X states in Section \ref{sec:LAQC}. We then apply this general result to determine the LAQC quantifier for the five families of one-parameter X states analyzed in this work. Then, in Section \ref{sec:LAQCswap},  we present the general result of applying the QCS scheme to X states and study the particular examples of one-parameter X states selected. Finally, we present some conclusions in Section \ref{sec:Conclusions}.

%%%%%%%%%%%%%%%%%%%%%%%%%%%%%%%%%%%%%%%%%%%%%%
%%%%%%%%%%%%%%%%%%%%%%%%%%%%%%%%%%%%%%%%%%%%%%
%%%%%%%%%%%%%%  X-states  %%%%%%%%%%%%%%%%%%%%
%%%%%%%%%%%%%%%%%%%%%%%%%%%%%%%%%%%%%%%%%%%%%%
%%%%%%%%%%%%%%%%%%%%%%%%%%%%%%%%%%%%%%%%%%%%%%

\section{One-parameter 2-qubit X states}\label{sec:Xstates}

The widely-studied family of 2-qubit states known as X states \cite{EstadosX, Quesada-XStates} due to the shape of their density matrix
\begin{eqnarray}\label{eq:estadosX}
\rho^{AB}_X =\mqty(
   a & 0 & 0 & r\vb{e}^{i\chi} \\
   0 & b & s\vb{e}^{i\xi} & 0 \\
   0 & s\vb{e}^{-i\xi} & c& 0 \\
    r\vb{e}^{-i\chi}& 0 & 0 & d)
\end{eqnarray} 
is a representative class of 2-qubit states since any arbitrary $\rho_{AB}$ can be transformed into a $\rho^X$ while preserving its entanglement \cite{XStates-Entanglement, Hedemann-XStates}. Although the above density matrix has seven independent real parameters, Zhou et al. \cite{Zhou-CanonicalXstates} established that applying local unitary transformations allows one to remove the two phase parameters $\chi$ and $\xi$. Therefore, one can map the 7-parameter family defined in Eq. \eqref{eq:estadosX} into a 5-parameter one
\begin{eqnarray}\label{eq:estadosX-real}
\rho_X =\mqty(
   a & 0 & 0 & r \\
   0 & b & s & 0 \\
   0 & s & c& 0 \\
   r& 0 & 0 & d)
\end{eqnarray}
where
\begin{subequations}
\begin{align}
    &a,b,c,d\geq0,\\
    &a+b+c+d=1,\\
    &s\leq\sqrt{bc},\qq{and}r\leq\sqrt{ad},
\end{align}
\end{subequations}
so that $\rho^X$ is hermitian, semi-positive definite, and $\Tr\qty(\rho^X)=1$.

Since the well-known Pauli matrices,
\begin{equation}
    \sigma_1=\mqty(\pmat{1})\qc \sigma_2=\mqty(\pmat{2})\qc \sigma_3=\mqty(\pmat{3}),
\end{equation}
along with the identity matrix $\mathbbm{1}_2$, form a complete basis of $2\times2$ matrices, we can relay on them to represent any $\rho_{AB}$. This representation is the Fano form \cite{Fano1983} of 2-qubit states
\begin{equation}\label{eq:2-qubitFanoForm}
    \rho^{AB}=\frac{1}{4}\, \sum_{i,j=0}^{3}T_{ij}\sigma_i\otimes\sigma_j,
\end{equation}
where
\begin{equation}\label{eq:sigma0}
    \sigma_0=\mathbbm{1}_2=\mqty(\imat{2})
\end{equation}
and
\begin{equation}
    T_{ij}=\Tr\qty[\qty(\sigma_i\otimes\sigma_j)\rho^{AB}].
\end{equation}

One can use \eqref{eq:2-qubitFanoForm} to parametrize \eqref{eq:estadosX-real} using $\qty{T_{30}, T_{03}, T_{11}, T_{22}, T_{33}}$. Usually labeled as the Bloch parameters and denoted as $\qty{x_3,y_3,T_1,T_2,T_3}$ in the literature, these are related to the previous ones in \eqref{eq:estadosX-real} by
\begin{equation}\label{eq:Bloch_Xstates-abcdwz}
\begin{aligned}
    x_3&=a+b-c-d,\\
    y_3&=a-b+c-d,
\end{aligned}\qq{}
\begin{aligned}
    T_1&=2(s+r),\\
    T_2&=2(s-r),\\
    T_3&=a-b-c+d.
\end{aligned}
\end{equation}

Since \eqref{eq:2-qubitFanoForm} has to be semi-definite positive and Hermitian in order to behave as a proper density operator, these parameter are required to satisfy the following properties:

\begin{equation}\label{eq:Bloch_Xstates-prop}
\begin{aligned}
    a,b,c,d&\in\mathbbm{R},\\
    a+b+c+d&=1,\\
    -1\leq x_3, y_3, T_1, T_2, T_3&\leq 1,
\end{aligned}\qq{}
\begin{aligned}
    r^2\leq{bc},\\
     s^2\leq{ad}.\\
     \qty(T_1\pm{}T_2)^2\leq1.
\end{aligned}
\end{equation}

After this succinct review of 2-qubit X states and some of their properties, we proceed to present the five families of one-parameter bipartite states that we will discuss in order to illustrate the redistribution of LAQC via a QC swapping protocol.

The first subset this type of X states analyzed in this article is the highly symmetrical Werner states \cite{Werner}, defined as bipartite states invariant under local unitary transformations. That is, states $\rho_w$ that satisfy
\begin{equation}\label{eq:Werner-definition-U(2)}
    \rho_w = \qty(\mathbbm{U}_A\otimes\mathbbm{U}_B)\,\rho_w\,\qty(\mathbbm{U}_A\otimes\mathbbm{U}_B)^\dagger,
\end{equation}
where $\mathbbm{U}_A,\mathbbm{U}_B\in U(2)$. Such 2-qubit states can be written as
\begin{equation}\label{eq:rho-Werner-Psi-}
    \rho_w = z\dyad{\Psi^-} + \frac{1-z}{4}\,\mathbbm{1}_4,
\end{equation}
where $\ket{\psi^-}$ is one of the four maximally entangled 2-qubit states known as Bell states:
\begin{subequations}\label{eq:Bell_states}
\begin{align}
    \ket{\psi^{\pm}}&=\frac{1}{\sqrt{2\,}}\, \qty(\ket{01}\pm\ket{10}),\label{eq:PsiMasMenos}\\
    \ket{\phi^{\pm}}&=\frac{1}{\sqrt{2\,}}\, \qty(\ket{00}\pm\ket{11}).\label{eq:PhiMasMenos}
\end{align}
\end{subequations}
The Bloch parameters of $\rho_w$ \eqref{eq:rho-Werner-Psi-} are
\begin{equation}\label{eq:rho_w-Bloch}
    x_3=y_3=0\qq{and} T_1=T_2=T_3=-z,
\end{equation}.

In 2010, Al-Qasimi and James \cite{Qasimi2011-AlphaBeta-Xstates} introduced $\alpha$ and $\beta$ X states, which belong to the Bell-Diagonal subset. Like Werner states, these X states have maximally mixed marginals, that is, null local Bloch parameters. As the authors analyzed the relationship between discord, entanglement, and linear entropy for 2-qubit systems, they found that those states are related to upper and lower bounds for the quantum discord \cite{qDiscord-Olliver, qDiscord-Henderson} (QD) and entanglement of formation \cite{PhysRevLett1997-EoF} (EoF).

For $\alpha$-states, which are related to the upper bound, the corresponding density matrix is
\begin{equation}\label{eq:X-states-alpha}
    \rho_\alpha = \frac{1}{2}\,\mqty(
    \alpha & 0 &0 & \alpha\\
    0 & 1-\alpha & 0 &0\\
    0 & 0 & 1-\alpha &0\\
    \alpha & 0 &0 &\alpha
    ),
\end{equation}
where $0\leq\alpha\leq1$, which can be written \cite{Roa2014EntSwap-Xstates} as
\begin{equation}\label{eq:X-states-alpha-ket}
\begin{split}
    \rho_\alpha = &\alpha\dyad{\phi^+}\\
    &\qq{}+\frac{1-\alpha}{2}\, \qty(\dyad{\psi^+} +\dyad{\psi^-}).
\end{split}
\end{equation}
Its Bloch parameters are
\begin{equation}\label{eq:X-states-alpha-Bloch}
    x_3=y_3=0\qc T_1 = \alpha= -T_2\qc T_3 = 2\alpha-1.
\end{equation}

For $\beta$-states, related to the lower bound of quantum discord for a given EoF, the density matrix is
\begin{equation}\label{eq:X-states-beta}
    \rho_\beta = \frac{1}{2}\,\mqty(
    \beta & 0 &0 & \beta\\
    0 & 1-\beta & 1-\beta &0\\
    0 & 1-\beta & 1-\beta &0\\
    \beta & 0 &0 & \beta
    ),
\end{equation}
where $0\leq\beta\leq1$. The corresponding Bloch parameters are
\begin{equation}\label{eq:X-states-beta-Bloch}
    x_3=y_3=0\qc T_1 = 1\qc T_2 = 1-2\beta=- T_3,
\end{equation}
and the state can be written as \cite{Roa2014EntSwap-Xstates}
\begin{equation}\label{eq:X-states-beta-ket}
    \rho_\beta =  \beta\,\dyad{\phi^+} +\qty(1-\beta)\dyad{\psi^+}.
\end{equation}
From the above expression, it is straightforward to realize that, for $\beta=0$ and $\beta=1$, the resulting state is maximally entangled.

On the other hand, since one can consider 2-qubit Werner states as a statistical mixture of a maximally entangled state \eqref{eq:Bell_states} and a maximally mixed state, such a realization can be used to define other one-parameter 2-qubit states that are similar statistical mixtures. For instance, Verstraete and Verschelde \cite{Verstraete-MAF} used one-parameter mixed states of a maximally entangled state and a basis state. That is, states that are written as
\begin{equation}\label{eq:rho-VV-general}
    \rho_v= F\,\dyad{\psi_{ME}} + \qty(1-F)\dyad{ij},
\end{equation}
where $\psi_{ME}$ is a maximally-entangled state \eqref{eq:Bell_states} and $\ket{ij}$ is one of the basis states, with $i,j=0,1$. In this article, we will consider a particular case for $\rho_v$ with $\ket{\psi_{ME}}=\ket{\psi^-}$ and $i,j=0$ so that
\begin{equation}\label{eq:rho-VV-PsiMenos-00}
    \rho_v= F\,\dyad{\psi^-} + \qty(1-F)\dyad{00},
\end{equation}
whose Bloch parameters are
\begin{equation}\label{eq:rho_v-Bloch}
    \begin{aligned}
        &x_3=y_3=1-F,\\
        &T_1=T_2=-F,\\
        &T_3=1-2F.
    \end{aligned}
\end{equation}.

Finally, Munro et al. \cite{Munro2001Maximizing} defined a family of maximally entangled 2-qubit mixed states $\rho_{MEMS}$ that, for a given linear entropy, have the maximum amount of entanglement possible:
\begin{equation}\label{eq:rho-MEMS}
    \rho_{MEMS} = \mqty(\Gamma &0 &0 &\gamma/2\\
                        0& 1-2\Gamma &0 &0\\
                        0 &0 &0 &0\\
                        \gamma/2&0 &0 &\Gamma),
\end{equation}
where $\Gamma=\Gamma\qty(\gamma)$ is a function of the parameter $\gamma$ 
\begin{equation}\label{eq:Gamma-rho_MEMS}
    \Gamma = \left\{
    \begin{aligned}
        &1/3\qc 0\leq\gamma<2/3.\\
        &\gamma/2\qc 2/3\leq\gamma\leq1.
    \end{aligned}
    \right.
\end{equation}
The Bloch parameters for this state are
\begin{equation}\label{eq:rho_MEMS-Bloch}
    \begin{aligned}
        &x_3=-y_3=1-2\Gamma,\\
        &T_1=-T_2=\Gamma,\\
        &T_3=4\Gamma-1.
    \end{aligned}
\end{equation}.

It should be noticed that for $2/3\leq\gamma\leq1$, $\rho_{MEMS}$ \eqref{eq:rho-MEMS} is equivalent to $\rho_v$ \eqref{eq:rho-VV-general} with $\ket{\psi_{ME}}=\ket{\phi^+}$ and $\ket{ij}=\ket{10}$.

Having presented the families of one-parameter X states we are interested in analyzing, we continue to review the quantum correlation we are interested in studying in the context of a QCS scheme.

%%%%%%%%%%%%%%%%%%%%%%%%%%%%%%%%%%%%%%%%%%%%%%
%%%%%%%%%%%%%%%%%%%%%%%%%%%%%%%%%%%%%%%%%%%%%%
%%%%%%%%%%%%%%%%  LAQC  %%%%%%%%%%%%%%%%%%%%%%
%%%%%%%%%%%%%%%%%%%%%%%%%%%%%%%%%%%%%%%%%%%%%%
%%%%%%%%%%%%%%%%%%%%%%%%%%%%%%%%%%%%%%%%%%%%%%

\section{Local-available quantum correlations of 2-qubit X states}\label{sec:LAQC}

In 2015, Mundarain and Ladrón de Guevara introduced local-available quantum correlations (LAQC) \cite{LAQC}, a quantum correlation defined in terms of two mutually unbiased bases (MUB) \cite{Schwinger_MUB}. Those MUBs are, on the one hand, the optimal computational basis, defined as minimizing classical correlations, and, on the other hand, their complementary basis, given as the one that maximizes the LAQC quantifier. 

Classical states are diagonal on a local basis, so one can define a classical state related to a given $\rho^{AB}$ by choosing the original local computational basis to write it as
\begin{equation}\label{eq:ClassicalState-Pi_rhoAB}
    \Pi_{\rho^{AB}}=\sum_i R_i\dyad{i\,i},
\end{equation}
where
\begin{equation}
    R_i\equiv\ev{\rho^{AB}}{i\,i}.
\end{equation}
Since this is only the classical state in a particular local basis, one can generalize this concept using the computational basis that results from applying local $U(2)$ transformations to the original local computational bases. Given the two local computational bases, $\qty{\ket{0}^{(I)},\ket{1}^{(I)}}$, with $I=A,B$, one can define new ones by having the following unitary operators acting on them
\begin{equation}\label{eq:U(2)-Gen}
    \mathbbm{U}_I = \mqty[\cos\qty(\frac{\theta_I}{2}) & \sin\qty(\frac{\theta_I}{2})   \\
   \sin\qty(\frac{\theta_I}{2}) \vb{e}^{i\phi_I} & -\cos\qty(\frac{\theta_I}{2}) \vb{e}^{i\phi_I}]\;\;\in\;U(2),
\end{equation} 
where
\begin{equation}
    0\leq \theta_I\leq\pi\qc 0\leq\phi_I\leq2\pi.
\end{equation}
This parametrization leads to a set of classical states $\qty{ \Pi_{\rho^{AB}} \qty(\theta_I,\phi_I)}$ related to $\rho^{AB}$, given by
\begin{equation}\label{eq:ClassicalState-Pi_rhoAB-gen}
    \Pi_{\rho^{AB}}\qty(\theta_I,\phi_I)=\sum_i R_{i_A,i_B}\qty(\theta_I,\phi_I)\dyad{\widetilde{i\,i}},
\end{equation}
where
\begin{subequations}
    \begin{align}
    R_{i_A,i_B}\qty(\theta_I,\phi_I)&= \ev{\mathbbm{U}_{A\otimes{}B}^\dagger\rho^{AB}\mathbbm{U}_{A\otimes{}B}}{i\,i}\nonumber\\
    &=\ev{\rho^{AB}}{\widetilde{i\,i}},\\
    \mathbbm{U}_{A\otimes{}B} &=\mathbbm{U}_A\otimes\mathbbm{U}_B\qc \mathbbm{U}_{I}\in U(2).
\end{align}
\end{subequations}

One can write the mutual information of a given $\rho^{AB}$ as 
\begin{equation}\label{eq:Info_Mutua-Prob}
    I\qty(\rho^{AB}) =\sum_{i,j} R_{i_A,j_B}\log_2R_{i_A,j_B} -\sum_{i,I}R_{i_I}\log_2R_{i_I},
\end{equation}
\noindent{}where 
\begin{subequations}\label{eq:P_theta_phi}
\begin{align}
    R_{i_A,j_B} &= \ev{\rho^{AB}}{i_A\,j_B}\\
    R_{i_I} &= \ev{\rho^I}{i_I},
\end{align}
\end{subequations}
and $\rho^I$, with $I=A,B$, is the reduced operator corresponding to the $I$ subsystem. Therefore, for a classical state $\Pi_{\rho^{AB}}$ \eqref{eq:ClassicalState-Pi_rhoAB-gen}, the mutual information \eqref{eq:Info_Mutua-Prob} depends on the local parameters $\theta_I$ and $\phi_I$. With this in mind, one can define the optimal computational basis as minimizing Eq. \eqref{eq:Info_Mutua-Prob}, with $R_{ij}\qty(\theta_I,\phi_I)$ and $R_{i_I}\qty(\theta_I,\phi_I)$. The obtained minimized mutual information is the classical correlation quantifier,
\begin{equation}\label{eq:ClassicalCorrelationsMeasure}
    \mathcal{C}\qty(\rho^{AB})= \min_{\theta_I,\phi_I}\qty\Big{I\qty[\Pi_{\rho^{AB}}\qty(\theta_I,\phi_I)]}.
\end{equation}

Given the optimal computational basis, one must determine its complementary one. To do so, one starts by writing $\rho^{AB}$ in the newly found optimal computational basis
\begin{equation}
    \tilde{\rho}^{AB}= \mathbbm{U}_{A\otimes{}B}\;\rho^{AB}\;\mathbbm{U}_{A\otimes{}B}^\dagger.
\end{equation}
In the above expression, $\mathbbm{U}_{A\otimes{}B}= \mathbbm{U}_{A\otimes{}B}\qty(\theta_I,\phi_I)$ is given in terms of the angles $\theta_A,\theta_B,\phi_A$, and $\phi_B$ that relates the original computational basis with the optimal one. Either by considering the associated complex Hadamard matrix \cite{Bengtsson2007-MUB, DURT2010-MUB} or a basis defined by the eigenvectors of $\sigma_\perp$, that is, the Pauli matrix vector $\va*{\sigma}$ projected in a direction perpendicular to the $\sigma_{\vu{n}}$ that specifies the optimal computational basis, one has to define the complementary basis. To do so, one determines classical states in this new basis 
\begin{equation}\label{eq:ClassicalState-Pi-hat_rhoAB}
    \hat{\Pi}_{\rho^{AB}}\qty(\Phi_I)=\sum_i \hat{R}_{i_A,i_B}\qty(\Phi_A,\Phi_B)\dyad{\widehat{i\,i}},
\end{equation}
where
\begin{subequations}
    \begin{align}
    \hat{R}_{i_A,i_B}\qty(\Phi_A,\Phi_B)&= \ev{\hat{\mathbbm{U}}_{A\otimes{}B}^\dagger\,\rho^{AB}\,\hat{\mathbbm{U}}_{A\otimes{}B}}{\widetilde{i\,i}}\nonumber\\
    &=\ev{\rho^{AB}}{\widehat{i\,i}},\\
    \hat{\mathbbm{U}}_{I} &= \frac{1}{\sqrt{2\,}}\;
    \mqty(1 & 1   \\
   \vb{e}^{i\Phi_I} & -\vb{e}^{i\Phi_I})\;\;\in{}U(2),
\end{align}
\end{subequations}
with $0\leq\Phi_I\leq2\pi$ and $I=A,B$.

Analogous to determining the classical correlation measure \eqref{eq:ClassicalCorrelationsMeasure}, one calculates the quantum mutual information for this classical state using Eq. \eqref{eq:Info_Mutua-Prob}. Mundarain and Ladrón de Guevara defined the LAQC quantifier $\mathcal{L}(\rho^{AB})$ as the maximum of the above-defined mutual information:
\begin{equation}    \label{eq:LAQC-definition}
    \mathcal{L}\qty(\rho^{AB}) = \max_{\Phi_A,\Phi_B}\qty{I\qty[\tilde{\Pi}_{\rho^{AB}}\qty(\Phi_A,\Phi_B)] }.
\end{equation}

Contrary to other quantum correlations, such as quantum discord \cite{qDiscord-Olliver, qDiscord-Henderson}, where exact analytical expressions are only possible \cite{GirolamiAdesso-QD-Xstates, Huang-QD-Xstates-WorstCaseScenario} for a reduced set of 2-qubit states like Bell-Diagonal states \cite{Luo2008QD-BellDiagonal}, there is one such result for the LAQC quantifier for 2-qubit X states \cite{LAQC_Xstates-sym, LAQC_Xstates-no_sym}. By defining the following function in terms of the Bloch parameters and diagonal elements of $\rho_X$ \eqref{eq:estadosX-real},
\begin{align}\label{eq:g3}
    g_3 &=  \frac{1}{4} \qty(a\, \log_2a +b\,\log_2b + c\, \log_2c +d\, \log_2d) \nonumber
        \\
    & \qq{} -\frac{1}{2}\qty\big[u\qty(x_3)+u\qty(y_3)],        
\end{align}
where $a,b, c$, and $d$ are the diagonal elements of $\rho_X$, and $u(x)=(1+x)\log_2(1+x)+(1-x)\log_2(1-x)$, the LAQC quantifier for 2-qubit X states is
\begin{equation}\label{eq:EstadosX-LAQC}
    \mathcal{L}\qty(\rho_X) = \max\qty\Big{u(T_1),u(T_2),g_3(x_3,y_3,T_3)}.
\end{equation}

\subsection{Application to specific one-parameter X-states}

With this result, one can directly compute the LAQC quantifiers for the chosen one-parameter X states discussed in the previous section. For Werner states \eqref{eq:rho-Werner-Psi-}, one has that
\begin{equation}\label{eq:LAQC-Werner}
    \mathcal{L}\qty(\rho_w) = \frac{1+z}{2}\log_2\qty(1+z)+ \frac{1-z}{2}\log_2\qty(1-z).
\end{equation}
One can consider the concurrence \cite{Wooters_Concurrence} as a measure of entanglement. In such a case, we have the well-known result
\begin{equation}\label{eq:Werner-Conc}
    C(\rho_w)=\max\qty{0\qc\frac{1}{2}\,\qty(3z-1)}.
\end{equation}
\begin{figure}[ht!]
    \centering
    \includegraphics[width=0.75\linewidth]{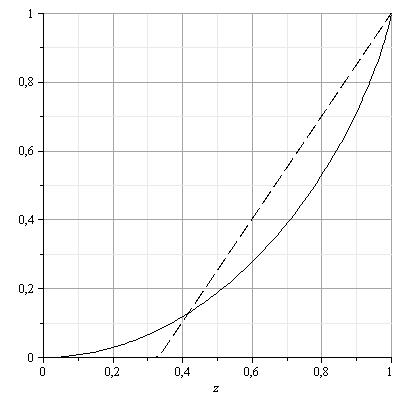}
    \caption{Concurrence (dashed) and LAQC (solid) for Werner states \eqref{eq:rho-Werner-Psi-}.}
    \label{fig:Werner-LAQC&Conc}
\end{figure}
We present the concurrence (dashed line) and LAQC quantifier (solid line) for Werner states in Figure \ref{fig:Werner-LAQC&Conc}.

As for the two other Bell Diagonal states included in this analysis, we first focus on $\alpha$-states. For them, using Eq. \eqref{eq:EstadosX-LAQC}, one can verify that the LAQC quantifier is
\begin{equation}\label{eq:LAQC-X-alpha}
    \mathcal{L}\qty(\rho_\alpha) = \frac{1+\alpha}{2}\log_2\qty(1+\alpha)+ \frac{1-\alpha}{2}\log_2\qty(1-\alpha).
\end{equation}
Regarding their concurrence, it is
\begin{equation}\label{eq:Conc-X-alpha}
    C(\rho_\alpha)=\max\qty{0,2\alpha -1}.
\end{equation}
For $\beta$-states, the LAQC quantifier has an analogue functional expression, given by
\begin{equation}\label{eq:LAQC-X-beta}
    \mathcal{L}\qty(\rho_\beta) = 1 + \beta\log_2\beta + \qty(1-\beta)\log_2\qty(1-\beta),
\end{equation}
while its concurrence is
\begin{equation}\label{eq:Conc-X-beta}
    C(\rho_\beta)=\max\qty{0,\qty|1-2\beta|}.
\end{equation}

\begin{figure}[ht!]
    \centering
    \includegraphics[width=0.75\linewidth]{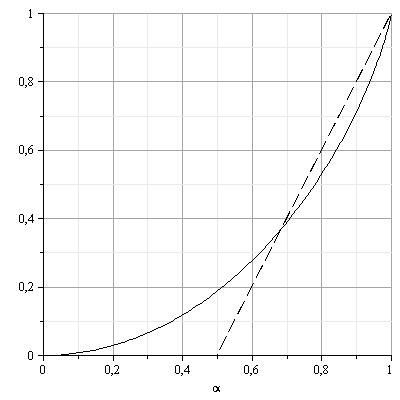} 
    \caption{Concurrence (dashed) and LAQC (solid) for $\alpha$-states \eqref{eq:X-states-alpha}.}
    \label{fig:X-alpha-LAQC&Conc}
\end{figure}

\begin{figure}[ht!]
    \centering
    \includegraphics[width=0.75\linewidth]{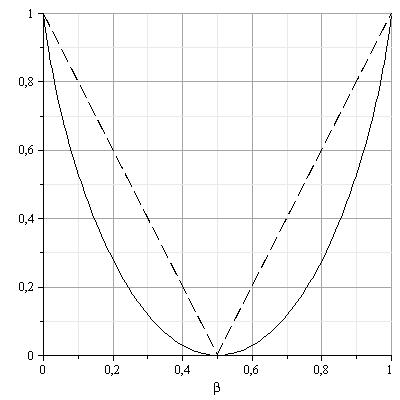}
    \caption{Concurrence (dashed) and LAQC (solid) for $\beta$-states \eqref{eq:X-states-beta}.}
    \label{fig:X-beta-LAQC&Conc}
\end{figure}
In Figure \ref{fig:X-alpha-LAQC&Conc}, we present the concurrence (dashed line) and LAQC quantifier (solid line) for $\alpha$-states \eqref{eq:X-states-alpha} while Figure \ref{fig:X-beta-LAQC&Conc} shows these quantifiers for $\beta$-states \eqref{eq:X-states-beta}. One can readily notice that, on the one hand, $\alpha$-states are separable for $0\leq\alpha\leq1/2$ and LAQC are higher than concurrence for up to $\alpha\sim0.6872$, after which the concurrence is bigger than LAQC. On the other hand, for $\beta$-states the concurrence is always above the LAQC quantifier, as is the case for the following two one-parameter X states analyzed in this study. Moreover, unlike all other states analyzed, $\beta$-states are only separable for $\beta=1/2$ and become maximally entangled states for $\beta=0$, where $\rho_\beta=\dyad{\psi^+}$, and $\beta=1$, where $\rho_\beta=\dyad{\phi^+}$, as can be readily seen in Eq. \eqref{eq:X-states-beta-ket}.

For $\rho_v$ \eqref{eq:rho-VV-general}, the LAQC quantifier has an equivalent expression to the one found for Werner states,
\begin{equation}\label{eq:LAQC-rhoVV}
    \mathcal{L}\qty(\rho_v) = \frac{1+F}{2}\log_2\qty(1+F)+ \frac{1-F}{2}\log_2\qty(1-F).
\end{equation}
As for the concurrence of these states, a simple calculation leads to
\begin{equation}\label{eq:rhoVV-Conc}
    C(\rho_v)=F.
\end{equation}

\begin{figure}[ht!]
    \centering
    \includegraphics[width=0.75\linewidth]{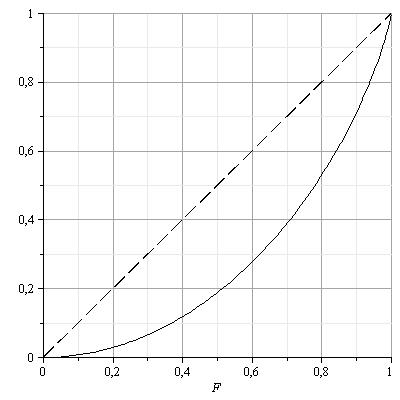}
    \caption{Concurrence (dashed) and LAQC (solid) for $\rho_v$ \eqref{eq:rho-VV-general}.}
    \label{fig:rhoVV-LAQC&Conc}
\end{figure}

In Figure \ref{fig:rhoVV-LAQC&Conc} we present the concurrence (dashed line) and LAQC quantifier (solid line) for $\rho_v$.

As for $\rho_{MEMS}$ \eqref{eq:rho-MEMS}, the results are analogous to the previous ones, only exchanging the state's parameter $F$ from Eq. \eqref{eq:rho-VV-general} to $\gamma$. That is,
\begin{subequations}
    \begin{align}
        \mathcal{L}\qty(\rho_{MEMS}) &= \frac{1+\gamma}{2}\log_2\qty(1+\gamma)\nonumber\\ &\qq{  } + \frac{1-\gamma}{2}\log_2\qty(1-\gamma),\label{eq:LAQC-rhoMEMS}\\
        C(\rho_{MEMS}) &=\gamma.
    \end{align}
\end{subequations}

In all of the above examples, we have that for at least a range of the state parameters, the concurrence is larger than the measure for LAQC. Even more so, only Werner and $\alpha$-states present values for which the LAQC measure is larger than concurrence. Nevertheless, the dynamics of this quantum correlation are more robust under Markovian noise than entanglement. While some states exhibit the so-called entanglement sudden death \cite{EntSuddenDeath}, this was not the case for LAQC \cite{LAQC_BD, LAQC_Xstates-sym, LAQC_Xstates-no_sym}. On the contrary, in all of the dynamics studied so far, there has only been an asymptotic death of this quantum correlation.

Moreover, analogous to what occurs for quantum discord and other quantum correlations beyond entanglement, LAQC only becomes zero if the state evolves into a classical one, that is, if it is diagonal in a local basis \cite{LAQC}. Therefore, despite the initial states possibly having less LAQC than entanglement, it is interesting to study how this quantum correlation behaves in a redistribution protocol such as the QC swapping one.

%%%%%%%%%%%%%%%%%%%%%%%%%%%%%%%%%%%%%%%%%%%%%%
%%%%%%%%%%%%%%%%%%%%%%%%%%%%%%%%%%%%%%%%%%%%%%
%%%%%%%%%%%%%% QC Swapping %%%%%%%%%%%%%%%%%%%
%%%%%%%%%%%%%%%%%%%%%%%%%%%%%%%%%%%%%%%%%%%%%%
%%%%%%%%%%%%%%%%%%%%%%%%%%%%%%%%%%%%%%%%%%%%%%

\section{LAQC swapping in 2-qubit X states}\label{sec:LAQCswap}

The starting point is considering two 2-qubit X states, labeled as $\rho_X^{AB}$ and $\rho_X^{CD}$, with Bloch parameters $\qty{x_3^{AB},y_3^{AB},T_1^{AB},T_2^{AB},T_3^{AB}}$ and $\qty{x_3^{CD},y_3^{CD},T_1^{CD},T_2^{CD},T_3^{CD}}$, respectively. We then perform a projective measurement on the $BC$ subsystems of $\rho^{ABCD}=\rho_X^{AB}\otimes\rho_X^{CD}$, using the pure state
\begin{equation}\label{eq:Phi_BC}
     \ket{\varphi}= \cos\qty(\frac{\xi}{2})\ket{00}^{BC} +\sin\qty(\frac{\xi}{2})\ket{11}^{BC},
\end{equation}
where $-\pi\leq\,\xi\,\leq \pi$. One can readily notice that for $\xi=\pm\pi/2$, $\ket{\varphi}=\ket{\phi^\pm}$ \eqref{eq:PhiMasMenos}, and that a simple unitary transformation $\mathbbm{U}=\mathbbm{1}\otimes\sigma_2$ applied to $\ket{\varphi}$ leads to an analogous state relating in the same manner to $\ket{\psi^\pm}$ \eqref{eq:PsiMasMenos}.

This projective measurement leads to $\rho_X^{AD}$, which is also of the X form \eqref{eq:estadosX-real}. The resulting state $\rho_X^{AD}$, before normalization, has the following Bloch parameters:
\begin{subequations}\label{eq:rhoAD-phi1-Bloch}
    \begin{align}
        x_3^{AD} =\; & \qty(1-x_3^{CD}\cos\xi)x_3^{AB} \nonumber
            \\ 
            &\qq{} + \qty(\cos\xi-x_3^{CD})T_3^{\,AB},\label{eq:rhoAD-phi1-x3}
        \\
        y_3^{AD} =\; & \qty(y_3^{AB}y_3^{CD}-T_3^{\,CD}) \cos\xi\nonumber
            \\
            & \qq{}+\qty(y_3^{CD}-y_3^{AB}T_3^{\,CD})\label{eq:rhoAD-phi1-y3},
        \\
        T_1^{AD} =\; & T_1^{\,AB}T_1^{\,CD}\sin\xi,\label{eq:rhoAD-phi1-T1}
        \\
        T_2^{AD} =\; & T_2^{\,AB}T_2^{\,CD}\sin\xi,\label{eq:rhoAD-phi1-T2}
        \\
        T_3^{AD} =\; & \qty(y_3^{CD}\cos\xi-T_3^{\,CD})T_3^{\,AB}\nonumber\\
            & \qq{} +\qty(y_3^{CD}-T_3^{\,CD}\cos\xi) x_3^{AB}.\label{eq:rhoAD-phi1-T3}
    \end{align}
\end{subequations}
The normalization factor is
\begin{equation}\label{eq:Norm-rhoX_AD}
    \mathcal{N}_{\rho_X^{AD}} = 1+y_3^{AB}x_3^{CD} +\qty(y_3^{AB}+x_3^{CD})\cos\xi,
\end{equation}
and the normalized Bloch parameters are therefore $T_{ij}=T_{ij}^{AD}/\mathcal{N}_{\rho_X^{AD}}$.

Given that $\rho_X^{AD}$ obtained as a result of applying the measurement in the QC swapping protocol is also an X state, we can use Eq. \eqref{eq:EstadosX-LAQC} to compute the LAQC measure. For general 2-qubit X states, the five independent parameters of each of the initial states, along with the free parameter $\xi$ of the pure state \eqref{eq:Phi_BC}, would seem to make it difficult to extract relevant information for this problem. Nevertheless, the definition and properties of LAQC established in \cite{LAQC} allow us to highlight relevant results. 

For instance, assuming that the QC swapping protocol is effective, that is, that the pure state $\ket{\psi}$ is entangled, we can assure the conditions under which the resulting state will be non-classical.

\begin{theorem}
    The resulting state $\rho_X^{AD}$ of a quantum correlation swapping protocol has non-null LAQC if the initial $\rho_X^{AB}$ and $\rho_X^{CD}$ have non-null $T_1$ and $T_2$ parameters.
\end{theorem}
\begin{proof}
    If $T_{1}\neq0$ and $T_{2}\neq0$ for both initial states, the resulting $\rho_X^{AD}$ is non-diagonal (that is, non-classical) since $T_1^{AD}$ \eqref{eq:rhoAD-phi1-T1} and $T_2^{AD}$ \eqref{eq:rhoAD-phi1-T2} will be non-zero. This characteristic, in turn, implies that at least two defining functions of the LAQC measure \eqref{eq:EstadosX-LAQC} for X states are not null.
\end{proof}

\begin{theorem}
    Given two non-classical initial states $\rho_X^{AB}$ and $\rho_X^{CD}$, the state $\rho_X^{AD}$ can only be classical if one initial state has $T_1=0$ and $T_2\neq0$ while the other has $T_1\neq0$ and $T_2=0$.
\end{theorem}
\begin{proof}
    Since $T_1$ and $T_2$ are the Bloch parameters involved in the coherences of an X state, as can be seen from Eq. \eqref{eq:Bloch_Xstates-abcdwz}, we have that if the initial states are non-classical, then either $T_1\neq0$ or $T_2\neq0$. Given that $T_1^{AD}$ depends on $T_1^{AB}$ and $T_1^{CD}$ while $T_2^{AD}$ depends on $T_2^{AB}$ and $T_2^{CD}$, the only possible combination for non-classical states resulting in a classical one in an effective QCS scheme is that $T_1^{AB}=T_2^{CD}=0$ or $T_2^{AB}=T_1^{CD}=0$.
\end{proof}

In what follows, we study five families of one-parameter 2-qubit states to better illustrate the behavior of LAQC in a QC swapping protocol.

%%%%%%%%%%%%%%%%%%%%%%%%%%%%%%%%%%%%%%%%%%%%%%
%%%%%%%%%%%%%%%%%%%%%%%%%%%%%%%%%%%%%%%%%%%%%%
%%%%%%%%%%%%%%% Análisis %%%%%%%%%%%%%%%%%%%%%
%%%%%%%%%%%%%%%%%%%%%%%%%%%%%%%%%%%%%%%%%%%%%%
%%%%%%%%%%%%%%%%%%%%%%%%%%%%%%%%%%%%%%%%%%%%%%
\subsection{Werner states}

We start by considering Werner states \eqref{eq:rho-Werner-Psi-} as initial states, with $z^{AB}$ and $z^{CD}$ as the corresponding state parameters. A simple calculation leads to the following Bloch parameters for the resulting $\rho_w^{AD}$:
\begin{subequations}
\begin{align}
    x_3 &= - z^{AB}\cos\xi,\\ 
    y_3 &= - z^{CD}\cos\xi,\\
    T_1 &=z^{AB}z^{CD}\sin\xi,\\
    T_2 &=-z^{AB}z^{CD}\sin\xi,\\
    T_3 &= z^{AB}z^{CD}.
\end{align}
\end{subequations}
With these parameters, we can directly determine the LAQC quantifier, given by
\begin{equation}\label{eq:rhoW-AD-LAQC}
    \begin{aligned}
        \mathcal{L}\qty(\rho_w^{AD}) = \;& \frac{1}{2} \qty(1+z^{AB}z^{CD}\sin\xi)
            \\ &\qq{} \times\log_2\qty(1+z^{AB}z^{CD}\sin\xi)
        \\
        & \;+ \frac{1}{2}\qty(1-z^{AB}z^{CD}\sin\xi)
            \\ &\qq{ } \times\log_2\qty(1-z^{AB}z^{CD}\sin\xi),
    \end{aligned}
\end{equation}
and the Concurrence,
\begin{equation}\label{eq:rhoW-AD-Conc}
\begin{aligned}
    C\qty(\rho_w^{AD}) =\;& \max\Bigg\{0, z^{AB}z^{CD}\qty|\sin\xi| \\
    & \qq{\hspace{5mm}}-\frac{1}{2}\Big[\qty(1-{z^{AB}}^2)\qty(1-{z^{CD}}^2)\\
    &\qq{\hspace{9mm}}+\qty(z^{AB}-z^{CD})^2\sin^2\xi\Big]^{\frac{1}{2}}\Bigg\}.
\end{aligned}
\end{equation}

\begin{figure}[t]
    \centering
    \includegraphics[width=0.48\linewidth]{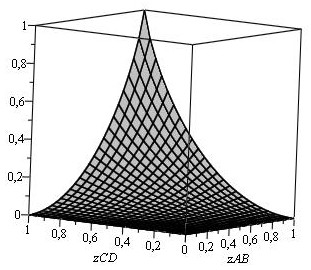}\hfill \includegraphics[width=0.48\linewidth]{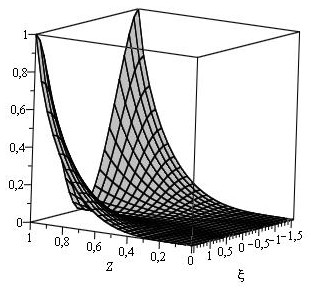}
    \caption{LAQC of the resulting state $\rho_w^{AD}$, considering $\xi=\pi/2$ (left) and $z^{AB}=z^{CD}=Z$ (right).}
    \label{fig:LAQC-WernerAD}
\end{figure}

In Figure \ref{fig:LAQC-WernerAD}, we present the graphs of the LAQC quantifier when $\xi=\pi/2$, that is, when $\ket{\varphi}=\ket{\phi^+}$, and when the initial Werner states have the same parameter, that is, when $z^{AB}=z^{CD}=Z$. As expected, when the projective measurement uses $\ket{\phi^+}$, there is only null LAQC for either $z^{AB}=0$ or $z^{CD}=0$. On the other hand, when considering equal state parameters, the LAQC quantifier is only null at all values of $Z$ for $\xi=0$, that is, for $\ket{\varphi}=\ket{00}$.

\begin{figure}[ht!]
    \centering
    \includegraphics[width=0.48\linewidth]{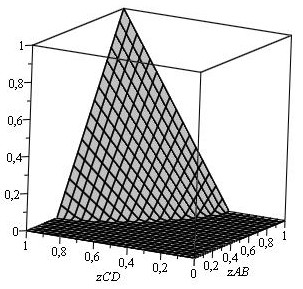}\hfill \includegraphics[width=0.48\linewidth]{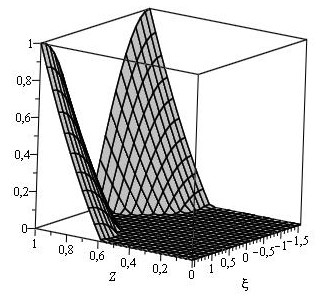}
    \caption{Concurrence of the resulting state $\rho_w^{AD}$, considering $\xi=\pi/2$ (left) and $z^{AB}=z^{CD}=Z$ (right).}
    \label{fig:Conc-WernerAD}
\end{figure}

\subsection{$\alpha$-states}

When we consider $\alpha$-states \eqref{eq:X-states-alpha}, with $\alpha^{AB}$ and $\alpha^{CD}$ the corresponding parameters, the resulting $\rho_\alpha^{AD}$ has the following Bloch parameters:
\begin{subequations}\label{eq:Alpha-AD-Bloch}
\begin{align}
    x_3=& \qty(1-2\alpha^{AB})\cos\xi,\\ 
    y_3=& \qty(1-2\alpha^{CD})\cos\xi,\\
    T_1=& \,\alpha^{AB}\alpha^{CD}\sin\xi,\\
    T_2=& \,\alpha^{AB}\alpha^{CD}\sin\xi,\\
    T_3 =& \qty(1-2\alpha^{AB})\qty(1-2\alpha^{CD}).
\end{align}
\end{subequations}
Given the above parametrization, the LAQC quantifier is analogous to the one previously obtained for Werner states. That is,
\begin{equation}\label{eq:Alpha-AD-LAQC}
    \begin{aligned}
        \mathcal{L}\qty(\rho_\alpha^{AD}) = \;& \frac{1}{2} \qty(1+\alpha^{AB}\alpha^{CD}\sin\xi)
            \\ &\qq{} \times\log_2\qty(1+\alpha^{AB}\alpha^{CD}\sin\xi)
        \\
        & \;+ \frac{1}{2}\qty(1-\alpha^{AB}\alpha^{CD}\sin\xi)
            \\ &\qq{ } \times\log_2\qty(1-\alpha^{AB}\alpha^{CD}\sin\xi).
    \end{aligned}
\end{equation}
Therefore, the surfaces presented in Figure \ref{fig:LAQC-WernerAD} also apply to $\alpha$-states. 

On the other hand, the resulting state $\rho_\alpha^{AD}$ is no longer entangled. After some algebraic manipulation, we can demonstrate that $C\qty(\rho_\alpha^{AD})=0$. This separability is not only the case when we use $\ket{\varphi}$ \eqref{eq:Phi_BC} as the pure state in the projective measurement, but also for $\ket{\varphi^\prime}=\mathbbm{X}^C \ket{\varphi}$. That is, the state resulting from applying the Pauli X-gate to $\ket{\varphi}$ \eqref{eq:Phi_BC}.

\subsection{$\beta$-states}

For $\beta$-states, with $\beta^{AB}$ and $\beta^{CD}$ denoting the corresponding state parameters, the resulting $\rho_{\beta}^{AD}$ can be characterized by the following Bloch parameters:

\begin{subequations}\label{eq:betaAD-Bloch}
\begin{align}
    x_3 &=\qty(1-2\beta^{AB})\cos\xi,\\ 
    y_3 &=(1-2\beta^{CD})\cos\xi,\\
    T_1 &= \sin\xi,\\ 
    T_2 &=-\qty(1-2\beta^{AB})\qty(1-2\beta^{CD})\sin\xi\\
    T_3 &= \qty(1-2\beta^{AB})\qty(1-2\beta^{CD}).
\end{align}
\end{subequations}
A direct calculation leads to the following LAQC quantifier \eqref{eq:EstadosX-LAQC}
\begin{equation}\label{eq:Beta-AD-LAQC}
        \mathcal{L}\qty(\rho_\beta^{AD}) = u\qty\Big[T_2\qty(\beta^{AB},\beta^{CD},\xi)],
\end{equation}
where we used the previously defined function $u(x)=(1+x)\log_2(1+x)+(1-x)\log_2(1-x)$.

\begin{figure}[ht!]
    \centering
    \includegraphics[width=0.48\linewidth]{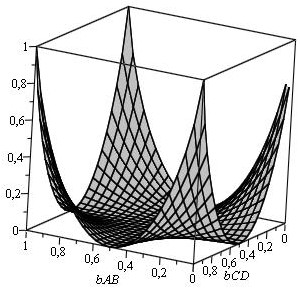}\hfill \includegraphics[width=0.48\linewidth]{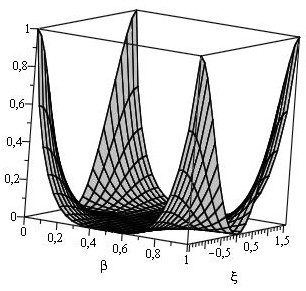}
    \caption{LAQC of the resulting state $\rho_\beta^{AD}$, considering $\xi=\pi/2$ (left) and $\beta^{AB}=\beta^{CD}=\beta$ (right).}
    \label{fig:LAQC-betaAD}
\end{figure}

As for the Concurrence,we have that, after some algebraic manipulation,
\begin{align}
    \mathcal{C}\qty(\rho_\beta^{AD}) =&\max\Bigg\{0,  \qty|\sin\xi|\nonumber\\
    &\times\qty|\qty(2\beta^{AB}-1)\beta^{CD}+1-\beta^{AB}|\\
    & -\Big[\qty(\beta^{AB}-\beta^{CD})^2\sin^2\xi\nonumber\\
    &\qq{}+4\beta^{AB\beta^{CD}}\qty(1-\beta^{AB})\qty(1-\beta^{CD})\Big]^{1/2}\Bigg\}.\nonumber
\end{align}

In Figure \ref{fig:LAQC-betaAD}, we present the graphs of the LAQC quantifier when $\xi=\pi/2$, that is, when $\ket{\varphi}=\ket{\phi^+}$, and when the initial $\beta$-states have the same parameter, that is, when $\beta^{AB}=\beta^{CD}=\beta$. As expected, when the projective measurement uses $\ket{\phi^+}$, there is only null LAQC for either $\beta^{AB}=1/2$ or $\beta^{CD}=1/2$, for which the corresponding initial state is separable (see Figure \ref{fig:X-beta-LAQC&Conc}). On the other hand, when considering equal state parameters, the LAQC quantifier is only null for an initial separable state, $\beta=1/2$, and for $\xi=0$, that is, for $\ket{\varphi}=\ket{00}$.

\begin{figure}[ht!]
    \centering
    \includegraphics[width=0.48\linewidth]{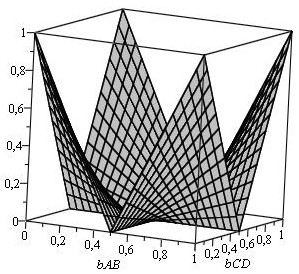}\hfill \includegraphics[width=0.48\linewidth]{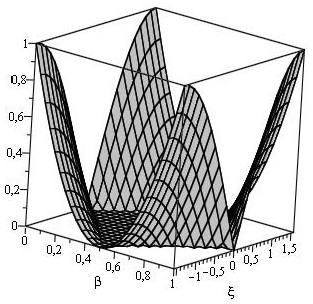}
    \caption{Concurrence of the resulting state $\rho_\beta^{AD}$, considering $\xi=\pi/2$ (left) and $\beta^{AB}=\beta^{CD}=\beta$ (right).}
    \label{fig:Conc-betaAD}
\end{figure}

\subsection{One-Parameter Mixed States involving a basis element}

By considering $\rho_v$ \eqref{eq:rho-VV-PsiMenos-00} as our initial states, with $F^{AB}$ and $F^{CD}$ characterizing each of them, the resulting $\rho^{AD}$ before normalization has the following Bloch parameters:

\begin{align}
    x_3 =\; & \qty[2-\qty(3-F^{CD})F^{AB} -F^{CD}]\qty(1+\cos\xi)\nonumber
        \\
        &\qq{} +F^{AB}F^{CD},\nonumber
    \\
    y_3 =\; &  \qty[2-\qty(3-F^{AB})F^{CD} -F^{AB}]\qty(1+\cos\xi)\nonumber
        \\
        &\qq{} +F^{AB}F^{CD},
    \\
    T_1 =\; & F^{AB}F^{CD}\sin\xi =-T_2,\nonumber
    \\
    T_3 =\;& \qty[2 - \qty(3-4F^{CD})F^{AB} -3F^{CD}]\qty(1+\cos\xi)\nonumber
        \\
        &\qq{} +F^{AB}F^{CD},\nonumber
\end{align}
with the normalization factor \eqref{eq:Norm-rhoX_AD} being
\begin{equation}\label{eq:Norm-rhoVV_AD}
\begin{aligned}
    \mathcal{N}_{\rho_v^{AD}} = &1 +\qty(1-F^{AB})\qty(1-F^{CD})\\
    &\qq{}+\qty\Big[2-\qty(F^{AB}+F^{CD})]\cos\xi.
\end{aligned}
\end{equation}

\begin{figure}[b]
    \centering
    \includegraphics[width=0.48\linewidth]{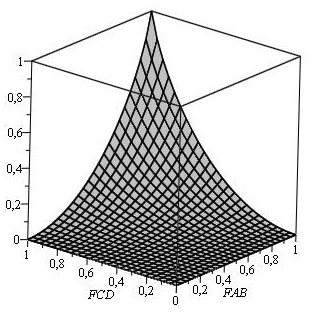}\hfill \includegraphics[width=0.48\linewidth]{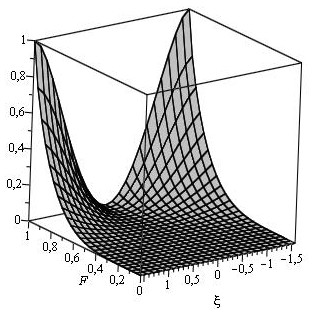}
    \caption{LAQC of the resulting state $\rho_v^{AD}$, considering $\xi=\pi/2$ (left) and $F^{AB}=F^{CD}=F$ (right).}
    \label{fig:LAQC-vvAD}
\end{figure}

From Eq. \eqref{eq:EstadosX-LAQC}, and after some calculations using the above-given parameters, the LAQC quantifier is
\begin{equation}\label{eq:rhoVV-AD-LAQC}
\begin{aligned}
    \mathcal{L}\qty(\rho_v^{AD}) =  &\frac{1+u_v}{2} \log_2\qty(1+u_v)\\
    &\qq{}+\frac{1-u_v}{2}\log_2\qty(1-u_v)
\end{aligned}
\end{equation}
where $u_v=T_1\qty(F^{AB},F^{CD},\xi)/\mathcal{N}_{\rho_v^{AD}}$.

Although the terms involved in the previous expression have a more complex functionality, given that the corresponding $\mathcal{N}_{\rho_X^{AD}}$ is no longer constant, the surfaces presented in Figure \ref{fig:LAQC-vvAD} are similar to the ones obtained for Werner and $\alpha$-states (see Figure \ref{fig:LAQC-WernerAD}).

Regarding the concurrence, a direct calculation leads to
\begin{align}\label{eq:rhoVV-AD-Conc}
    C\qty(\rho_v^{AD}) =\;& \mathcal{N}_{\rho_v^{AD}}^{-1}\max\qty\Big{0, C_1\qty(\rho_v^{AD})},\\
    C_1\qty(\rho_v^{AD}) =\;&
   \frac{1}{4}\,F^{AB}F^{CD}\qty|\sin\xi| - \qty(1-\cos\xi)   \nonumber
    \\
    &\;\times\sqrt{F^{AB}F^{CD}\qty(1-F^{AB})\qty(1-F^{CD})}.\nonumber
\end{align}

\begin{figure}[ht!]
    \centering
    \includegraphics[width=0.45\linewidth]{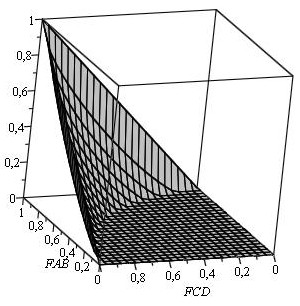}\hfill \includegraphics[width=0.51\linewidth]{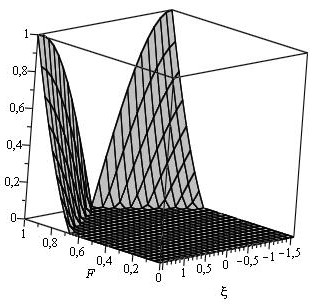}
    \caption{Concurrence of the resulting state $\rho_v^{AD}$, considering $\xi=\pi/2$ (left) and $F^{AB}=F^{CD}=F$ (right).}
    \label{fig:Conc-rhoVV_AD}
\end{figure}

In Figure \ref{fig:Conc-rhoVV_AD}, we present the surfaces corresponding to the Concurrence when $\ket{\varphi}=\ket{\phi^+}$, and when the initial $\rho_v$ states \eqref{eq:rho-VV-PsiMenos-00} have the same parameter, that is, when $F^{AB}=F^{CD}=F$. 

Unlike what we observed for the initial states, where the LAQC quantifier is always less than Concurrence (see Figure \ref{fig:rhoVV-LAQC&Conc}), the resulting state $\rho^{AD}$ is separable for a large sector of the initial parameters $F^{AB}$ and $F^{CD}$. On the other hand, the LAQC quantifier does not become zero unless one of the initial states is a basis state, that is, for $F^{AB}=0$ or $F^{CD}=0$.

\subsection{Maximally Entangled Mixed States} 

Finally, when we consider MEMS \eqref{eq:rho-MEMS} as our initial states, with $\gamma^{AB}$ and $\gamma^{CD}$ as their respective parameters, the resulting $\rho^{AD}$ before normalization has the following Bloch parameters
\begin{subequations}
\begin{align}
    x_3=\;&\qty[\qty(1+2\Gamma^{CD}) \Gamma^{AB} -\Gamma^{CD}] \qty(1+\cos\xi)\nonumber
        \\
        &\qq{ }+2\qty(1-3\Gamma^{AB})\Gamma^{CD},
    \\
    y_3=\;&\qty[\qty(1+2\Gamma^{AB})\Gamma^{CD} -\Gamma^{AB}] \qty(1+\cos\xi)\nonumber
        \\
        &\qq{ }-2\qty(1-\Gamma^{AB})\Gamma^{CD},
    \\
    T_1=\;& \frac{1}{2}\,\gamma^{AB}\gamma^{CD}\sin\xi,
    \\
    T_2=\;& \frac{1}{2}\,\gamma^{AB}\gamma^{CD}\sin\xi,
    \\
    T_3=\;& \qty(\Gamma^{CD}-\Gamma^{AB})\qty(1+\cos\xi)\nonumber
        \\
        &\qq{ }-2\qty(1-3\Gamma^{AB})\Gamma^{CD},
\end{align}
\end{subequations}
where $\Gamma^{AB}=\Gamma\qty(\gamma^{AB})$ and $\Gamma^{CD}=\Gamma\qty(\gamma^{CD})$ as defined in Eq. \eqref{eq:Gamma-rho_MEMS}. The normalization factor in this case is
\begin{equation}\label{eq:Norm-rhoMEMS_AD}
\begin{aligned}
    \mathcal{N}_{\rho_{MEMS}^{AD}} = &\qty(\Gamma^{AB}-\Gamma^{CD})(1+\cos\xi)\\
    &\qq{} +2\qty(1-\Gamma^{AB})\Gamma^{CD}.
\end{aligned}
\end{equation}

Again, the LAQC quantifier \eqref{eq:EstadosX-LAQC} has a similar expression to the previous cases. The maximization involved leads to $\mathcal{L}\qty(\rho_{MEMS}^{AD})= g_1\qty[\tilde{T}_1\qty(\gamma^{AB},\gamma^{CD},\cos\xi)]$, noticing that $\tilde{T}_1 = T_1\qty(\gamma^{AB},\gamma^{CD},\xi)/\mathcal{N}_{\rho_{MEMS}^{AD}}$. Therefore, the curves presented in Figure \ref{fig:LAQC-vvAD} also represent the graphical behavior of the LAQC quantifier for these states.

On the other hand, regarding Concurrence, the projective measurement involved in the quantum correlation swapping protocol using state \eqref{eq:Phi_BC} leads to a separable state. We can verify directly that $C\qty(\rho_v^{AD}) =0$ for all $0\leq\gamma^{AB},\gamma^{CD}\leq1$.
    
%%%%%%%%%%%%%%%%%%%%%%%%%%%%%%%%%%%%%%%%%%%%%%
%%%%%%%%%%%%%%%%%%%%%%%%%%%%%%%%%%%%%%%%%%%%%%
%%%%%%%%%%%%% Conclusiones %%%%%%%%%%%%%%%%%%%
%%%%%%%%%%%%%%%%%%%%%%%%%%%%%%%%%%%%%%%%%%%%%%
%%%%%%%%%%%%%%%%%%%%%%%%%%%%%%%%%%%%%%%%%%%%%%

\section{Summary and Conclusions}\label{sec:Conclusions}

We have successfully applied the standard quantum correlation swapping (QCS) scheme to the redistribution of local-available quantum correlations (LAQC). Using 2-qubit X states as initial states, we derived a general expression of the resulting state after performing a projective (von Neumann) measurement on two of the subsystems. Since this resulting state is also of the X type, we established conditions on the initial states so that the resulting state has a non-null LAQC.

As expected from previous studies of this quantum correlation, we can ensure that the resulting state of the QCS scheme will, in general, have a non-zero LAQC measure if the initial states are non-classical. It requires a specific combination of the Bloch parameters involved in the quantum coherences of the initial states to lead to a classical swapped state.

To illustrate this general result, we analyzed five families of one-parameter 2-qubit X states. Starting with the highly symmetrical Werner states, we also included two families of Bell-Diagonal states known as $\alpha$ and $\beta$ states. Furthermore, we included a one-parameter mixed state involving a basis element and maximally entangled mixed states as examples of X states with non-maximally mixed marginals. In this last category, we also analyzed the maximally entangled 2-qubit mixed states introduced by Munro et al. \cite{Munro2001Maximizing}. 

We compared our results for the LAQC quantifier with the redistribution of entanglement by analyzing the concurrence of the final state. As expected from previous results regarding this quantum correlation, the LAQC measure did not cancel for any of the studied cases if the initial states and the one used in the projective measurement were non-classical. 

In particular, we must highlight that, even though the initial states had their entanglement measure above the LAQC one for at least some parameter range, the resulting states have larger sections with non-zero LAQC while being separable. Moreover, for the one-parameter mixed state involving a basis element defined in Eq. \eqref{eq:rho-VV-PsiMenos-00}, whose entanglement measure is always above the LAQC one, we obtain a state that is separable for a wide range of the initial state parameters, while only having null LAQC for a particular set of initial conditions. 

The difference in the behavior of entanglement and LAQC in this QC swapping protocol is even more noticeable when analyzing the maximally entangled mixed states (MEMS) defined in Eq. \eqref{eq:rho-MEMS}. While we again have that the conditions for the final state $\rho_X^{AD}$ to have zero LAQC are very specific, it turns out to be always separable since its concurrence cancels out. TWe also observed this phenomenon for $\alpha$-states, for which the resulting concurrence is always zero. Those states are always separable independently of the adjustments to their initial parameters and therefore hinder the effectiveness of entanglement distribution with them.

The conditions that we established for general X states to have an effective LAQC redistribution, as well as the robustness under the QCS scheme shown by all studied states, open the possibility for this quantum correlation to be considered a genuine resource in quantum information technology.

%%%%%%%%%%%%%%%%%%%%%%%%%%%%%%%%%%%%%%%%%%%%%%
%%%%%%%%%%%%%%%%%%%%%%%%%%%%%%%%%%%%%%%%%%%%%%
%%%%%%%%%%% Agradecimientos %%%%%%%%%%%%%%%%%%
%%%%%%%%%%%%%%%%%%%%%%%%%%%%%%%%%%%%%%%%%%%%%%
%%%%%%%%%%%%%%%%%%%%%%%%%%%%%%%%%%%%%%%%%%%%%%

\section*{Acknowledgments}
The author would like to thank Prof. Bellorín (USB) for his comments and suggestions to the initial manuscript and acknowledge the support of the research group GID-30, \emph{Teoría de Campos y Óptica Cuántica}, at Universidad Simón Bolívar, Venezuela.

\section*{Declarations}

No funding was received to assist with the preparation of this manuscript.

\section*{Data availability statement}
The author declares that no datasets were generated or analyzed during this study. Therefore, no underlying data are available for this article.

%\newpage
\bibliographystyle{apsrev}
\bibliography{biblio-qit,biblio-qit2}
\include{biblio-qit,biblio-qit2}

\end{document}